Systematic Design and Evaluation of Social Determinants of Health Ontology (SDoHO)

Yifang Dang[1], Fang Li[1], Xinyue Hu[1], Vipina K. Keloth[1], Meng Zhang[1], Sunyang Fu[2], Jingcheng Du[1], J. Wilfred Fan[2], Muhammad F. Amith[1,3], Evan Yu[1], Hongfang Liu[2], Xiaoqian Jiang[1], Hua Xu[1], Cui Tao[1]*

[1]School of Biomedical Informatics, University of Texas Health Science Center at Houston, Houston, TX, USA

[2]Department of Artificial Intelligence and Informatics, Mayo Clinic, Rochester, MN, USA

[3]Department of Information Science, University of North Texas, Denton, TX, USA

*Address for correspondence:

Cui Tao, School of Biomedical Informatics, University of Texas Health Science Center at Houston, 7000 Fannin, Suite 600, Houston, TX 77030; Phone: 713-500-3981; Fax: 713-500-3929; e-mail: Cui.Tao@uth.tmc.edu

**Abstract**

Social determinants of health (SDoH) have a significant impact on health outcomes and well-being. Addressing SDoH is the key to reducing healthcare inequalities and transforming a "sick care" system into a "health promoting" system. To address the SDOH terminology gap and better embed relevant elements in advanced biomedical informatics, we propose an SDoH ontology (SDoHO), which represents fundamental SDoH factors and their relationships in a standardized and measurable way. The ontology formally models classes, relationships, and constraints based on multiple SDoH-related resources. Expert review and coverage evaluation, using clinical notes data and a national survey, showed satisfactory results. SDoHO could potentially play an essential role in providing a foundation for a comprehensive understanding of the associations between SDoH and health outcomes and providing a path toward health equity across populations.

**Introduction**

There is increasing awareness that medical care alone cannot improve population health if social, economic, and environmental issues are not addressed.[1] The non-medical factors that influence health outcomes are known as social determinants of health (SDoH). SDoH are "the conditions in the environments where people are born, live, learn, work, play, worship, and age that affect a wide range of health, functioning, and quality-of-life outcomes and risks."[2] SDoH are closely tied to health behaviors, lifestyle, and interpersonal relations,[3] and an increasing number of studies provide evidence for the impact of SDoH on health.[1] Notably, SDoH are estimated to account for between 30% and 55% of health outcomes.[4] Modifiable behaviors and community exposures are involved in the development of 70–90% of chronic diseases[5–7] and 60% of preventable deaths in the U.S.[8]

Furthermore, SDoH significantly affects health inequity. According to the World Health Organization, health and illness status follows a social gradient in countries at all income levels; i.e., people in lower socioeconomic positions tend to have worse health conditions.[4] In the United States, race/ethnicity associated with, to a large extent, an individual's disease risk, quality of care received, and prospects of wellness.[9] Another population-based cohort study from Canada demonstrated that individuals who live in the most unfavorable environments (i.e., materially resource-deprived areas, residentially unstable neighborhoods, and low-income neighborhoods) had fewer opportunities to receive a liver transplantation (with around a 40% reduced subhazard, $p < 0.01$) when diagnosed with decompensated cirrhosis or hepatocellular carcinoma.[10]

Although SDoH is a growing area of focus in healthcare and is becoming a key tool for addressing healthcare inequalities and disparities worldwide, currently, there are differences in its scope and domains among organizations. For example, the five categories classified by Healthy People 2030 include economic stability, education access and quality, healthcare access and quality, neighborhood and built environment, and social and community context.[11] The categories of Logical Observation Identifiers, Names, and Codes (LOINC) differ slightly and include economic stability, education, health care, neighborhood, social and community context, and food. Wild's framework includes concept of the exposome, specifically, the external exposome, which ranges from individual behaviors to environmental and broader societal aspects.[12] A related concept is the socio-behavioral determinants of health (SBDH), which concern social, behavioral, and environmental aspects of an individual's life that affect individual and community health.[13] Although researchers study the relationship of SDoH to various illnesses,[14–16] their focus generally is on a specific disease domain, and the categorization of the factors are not uniform across studies. These heterogeneous categorizations of SDoH can impede our understanding of the associations between the factors and health conditions. Thus, there is a need to standardize the categorization of SDoH.

Ontologies, which are a shared vocabulary in a specific domain, are widely used tools to identify, manage, and share semantic knowledge in complex areas.[17] Ontologies can assist with knowledge management and reasoning to improve semantic interoperability across systems or multiple data sources. In addition, ontologies can be used to test the consistency and ensure data quality, as they explicitly define the data types and precise terms.[18] Furthermore, ontologies enhance computational power by reducing semantic disambiguation in deductive inferences and enable complex, logical assertions and queries.[19] Extending ontologies to include the use of artificial intelligence can provide knowledge for computers to help with decision support. As ontologies provide interoperability and formal definitions of the terms and structure of the domain and subdomain relationship, an ontology-based approach can address the issue of heterogeneity.[20]

Currently, there are ontologies/terminologies that cover certain aspects of SDoH. As summarized in Table 1, the Ontology of Medically Related Social Entities (OMRSE), initiated in 2011 and developed on an ongoing basis, was created based on topics relevant to health that are related to societal roles.[21,22] For example, in their early framework, the OMRSE group defined the representation of demographic information, including gender and marital status. In their recent development, the group added representations of other aspects, such as organizational roles, healthcare provider roles, and enrollment in an insurance plan. OMRSE notes, however, that their model is not a comprehensive representation of social aspects that can influence human health.

[Insert Table 1 about here]

Melton et al. focused on information models designed based on public health surveys and clinical social history text and covered alcohol, tobacco, and drug use, and occupation.[23] Their study compared models constructed from public health and clinical sources and concluded that the models were similar, with the exception that the public health survey model provided a more complex view of the clinical model. Melton's model, however, did not capture a sufficient number of factors in the social environment, such as neighborhood or societal and community context, that could affect human health. The Semantic Mining of Activity, Social, and Health (SMASH) data system ontology, first introduced in 2017, describes health social networks, including the interrelations among health, specifically weight-related states or conditions, social activities, and daily physical activities.[24,25] The SMASH ontology, however, focuses on only these interrelations and lacks many of the aspects of SDoH, including economics, education, and the healthcare system. Gharebaghi et al. introduced an ontology that conceptualizes the social and environmental aspects of people with motor disabilities (PWMD).[26] It is aimed at being assistive for PWMD in terms of socio-environmental dimensions but was not developed to be generalizable for populations with other conditions. Kim et al. developed the Physical Activity Ontology (PACO) to address the heterogeneity of physical activity descriptions.[27] Their concern was physical or social-physical activities that influence health status, and they did not cover other aspects of community influence on human health, such as economics, food, or education. For their Social Interventions Research & Evaluation Network (SIREN), Arons et al. collected SDoH-related codes from several common medical coding systems. The network links factors from 20 SDoH domains to standardized medical systems but lacks comprehensiveness in terms of other SDoH factors and the relationships among them. In the ontology driven framework proposed by Rousseau et al., authors collected a list of SDoH-related concepts, however, the semantic meaning or relations were not defined in this ontology[28]. Lack of more than one hierarchy can also be a disadvantage of future application of the work.

Standardizing the SDoH factors with an ontology approach can address the challenge of heterogeneity embedded in SDoH definitions, categorizations, and applications. As noted, most works lack a comprehensive set of SDoH factors and measurements that researchers and practitioners can apply in the medical and public health fields. Therefore, we propose an ontology (SDoHO) that aims to comprehensively represent the concepts, hierarchies, and relations pertinent to SDoH factors. Well-collected measurements make our proposed ontology applicable to downstream applications, including clinical medicine, public health, and biomedical informatics, facilitating systematic SDoH knowledge representation, integration, and reasoning.

## Results

SDoHO is a comprehensive ontology with well-defined classes (concepts) and properties (relationships and features) represented in the Web Ontology Language (OWL2).[29] The current version has 706 classes, 105 object properties, and 20 data properties, with 1,542 logical axioms and 966 declaration axioms. Figure 1 shows its core conceptual framework, including the main top-level classes and some subclasses to demonstrate the usage of object properties with partial details of the framework. The main top-level hierarchy comprises nine categories, among which five are adapted from Healthy People 2030 (i.e., economic stability, education, health care, neighborhood, and social and community context). One of these categories is from the LOINC (i.e., food), and three are defined based on other resources newly added (i.e., SBDH[13] for behavior and lifestyle, OMRSE[21,22] for demographic, and PhenX Toolkit[30] for measurement). The maximum depth of the class hierarchy in the proposed SDoHO is six.

[Insert Figure 1 about here]

## Class

SDoH factors and related information were collected and defined as in OWL classes, resulting in a total of 706 classes in the SDoHO. There are nine top-level classes that represent the main SDoH related factors. As shown in Figure 2, the nine main top-level classes are "Element_Relevant_to_Behavior_and_Lifestyle," "Demographic," "Element_Relevant_to_Education," "Element_Relevant_to_Social_and_Community_Context," "Element_Relevant_to_Health_Care," "Element_Relevant_to_Economic_Stability," "Element_Relevant_to_Neighborhood," "Element_Relevant_to_Food," and "Measure_and_Index_and_Score." Six of the nine classes were drawn from Healthy People 2030 and LOINC.[31] "Element_Relevant_to_Behavior_and_Lifestyle" was included after reviewing the SBDH concept.[13] "Demographic" was added to represent the subject individual's societal role.[21,22] "Measure_and_Index_and_Score" was created, whereas measurement information of the other major top levels was collected. Under the "Measure_and_Index_and_Score," there are 21 qualitative and quantitative measurements collected and linked to the relevant factors. Among the 21 measurements, four were self-defined to accommodate the measurements with flexibility, such as "adherence," which has permissible values of "good_adherence" and "poor_adherence." Two terms "Exercise_Effect" and "Modifier" were adopted from the PACO. We expanded the "Modifier" class to measure behaviors other than physical activity. Fifteen are survey questions collected from various sources to measure SDoH factors, such as physical activity, vital signs, alcohol drinking, smoking, air quality, and healthy food accessibility. We imported and organically reused elements from some existing ontologies, including Simple Knowledge Organization System (SKOS), Time event ontology (TEO), and PACO, to formally represent aspects of certain professions. The purpose of reusing existing ontologies is to ensure standardization and interoperability while representing relations among SDoH factors with versatility. PACO, for example, can semantically explain major fields that involve physical activities.[27] In addition, importing TEO extended the dimensions of SDoHO to represent time lapses of the events.[32] We also defined a "Person" class to demonstrate how the defined classes and properties can be used to represent SDoH information for each individual. Class definitions are semantically defined appropriate classes, using OWL axioms. The current ontology includes 21 formally defined classes that can enable semantic inference and automatic classification.

[Insert Figure 2 about here]

**Property**

We defined 105 object properties, 20 data properties, and 31 annotation properties to represent relationships between SDoH factors and people/patients as well as among the factors. Of the 105 object properties, 23 were used to describe the relations between people/patients and SDoH factors. For example, for "Person" "has_race," "Race" describes the relation of "Person" and "Race." For the interrelationships among factors, the main object property is "relates"; for example, "Occupation" "relates" some "Exposure_to_Carcinogen_and_Pathogen" was used to show the relationship between occupations and possible work-related environmental exposure. Another major object property between SDoH factors is "has_measure," which links the SDoH factors to the relevant measurements under the "Measure_and_Index_and_Score" class. This object property was designed to retrieve the quantitative and qualitative information relevant to the corresponding factor. For instance, the SDoH factor "Food_Swamp" "has_measure" with some "Traditional_Retail_Food_Environment_Index" (RFEI), which measures the ratio of healthy and unhealthy food options in a range of geographical radius (Figure 3). With the information on the address of the neighborhood, this ontology can facilitate the RFEI score calculation and further evaluate patients' accessibility to healthy food options. The equation can be further transformed into a computational query, which can be practically efficient when encountering patients, such as those with obesity conditions, who need nutritious access. Further, we added the object property of "has_time_flag," which can represent time properties to the chronicle state of a subject or event. With richly defined object properties, we can also express n-ary relationships when more information is added to the statement triple. Our primary solution to the n-ary relationship is to define an additional attribute that describes the relation.[33] A new node that can represent the relation instance itself was inserted between the original subject and the original objects. The new added node describes the original relation as an instance, and other participants in the original objects can describe the additional information related to the relation. To semantically represent the patient's smoking history in the past, smoking cessation, and current non-smoking status, new attributes were inserted to describe the complex n-ary semantic relations (please see the Use Case section).

[Insert Figure 3 about here]

In addition to the built-in data property, we defined and organically imported 19 data properties to represent the relationship between concepts and data. We defined "has_number" in considering future applications and measurements. The use case example in Figure 4a provides a demonstration of the data property "has_number," which is defined as the numeric values of any measures. Further, 12 data properties imported from TEO assisted with classifying the time-related relation between concepts and data, such as "TEO: hasAgeValue," as shown in Figure 4b. In the example, this data property recorded and normalized the textual age into a float. PACO had one data property, "PACO: hasTotalAmountMin," that was used in describing the total number of minutes in the relations.

[Insert Figure 4 about here]

Further, we utilized annotation properties in the proposed ontology. In addition to the built-in annotation properties with predefined semantics in OWL2 and RDFS, there were four relevant ones adopted from TEO and ten imported from SKOS. Annotation properties from SKOS assisted in constructing the ontology basic needs, even in other ontologies; "definition (skos:definition)" and "alternative label" were utilized mainly in the imported ontologies, such as TEO. The annotation properties from TEO were used mainly in the TEO specific situation. With the utilization of object properties, data properties, and annotation properties, the proposed ontology has comprehensively standardized the representation of features and relationships of the classes.

## Use Case Demonstration

As noted, the design of the SDoHO aims to address the challenge of heterogeneity in SDoH research. We constructed two use cases with simulated data to represent the potential usage of the proposed SDoH ontology to represent patient information for different disease scenarios and extracted from different sources. The first case is used to illustrate how to represent data from clinical notes of a psychiatric-disorder patient. The second case is used to present a simulated survey note constructed to describe a patient diagnosed with HIV infection. The following are the notes for the first synthetic patient and represented in Figure 4a. "The patient graduated from high school. She is retired and worked as a nurse in the past. She does not smoke or use alcohol. However, the patient used tobacco but stopped. Her parents ignored her when she was growing up. She lives alone after divorce. She has a personal loan of 30k."

The text of the notes can be represented in RDF triples with respect to the SDoHO to enable semantic queries, inference, consistency checking, and minimize ambiguity. As shown in Figure 4, *case a* represents the first-case patient's data storage in graphed RDF triples. For example, the text, "She is retired, and worked as a nurse in the past," involves two chronicle statuses, current retirement, and past occupation. Therefore we used two chains of triples to represent the information. The first chain of triples can be "patient_a" "has_employment_status" "retired"; and "retired" "has_time_flag" "current" to describe the patient's current employment status (retirement). Then, to describe the patient's occupation before her retirement, the second chain of triples can be "patient_a" "has_occupation" "nurse"; and "nurse" "has_time_flag" "past." This patient's current employment and past occupation can be queried directly with the clear time-flag separation. This time flag can also be used in other triples to describe the chronicle status of a class instance with flexibility. In addition, in a more complex sentence, such as the *case a* text, "She does not smoke or use alcohol. However, the patient used tobacco but stopped." The short text involved not only the chronicle status of the instance but also the negation of the object property and change of the behavior state. As shown in Figure 4, we represent the n-ary relation text in three chains of triples for each substance use behavior. The two current negative substance use behaviors (substance_use_429 and substance_use_430) were represented by the negation assertion, and one changed substance behavior (substance_use_431), by a historical time flag and a change of behavior instance. For each specific behavior, the n-ary relationship was represented by two chains of object properties, first, to state the detailed behavior, smoking or alcohol drinking in this example, and, second, to state the chronicle status. In short, information from this psychiatric clinical note was fully represented with versatile functions of the proposed ontology.

The second example shows survey answers from a synthetic HIV-infected patient. The notes are as follows: "The patient is 34 yo black male, who was diagnosed with HIV last month. He answered that he is homosexual that had about 3 male sex partners in the past year. Not always use condom. He was in prison in the past but bailed out."

In this example case, *patient b* was represented with respect to the ontology in Figures 4b and 5. Figure 4b shows the conceptual level representation, and Figure 5 shows the ontology representation in OWL. Each textual information from the synthetic survey answer was stored in the ontology. Other than the triples, the proposed ontology has versatile functions that can make inference queries, represent n-ary relationships by other means, and standardize numeric value. Inference queries, such as the HIV patient's sexual behavior of "Men_Who_Have_Sex_with_Men" (MSM), can be achieved by the query design, as detailed in Figure 5, example *a*. MSM was designed to be equivalent to a male person who identified as homosexual or bisexual. Therefore, the reasoner (HermiT 1.4.3.456) can infer that the *case b* patient from Figure 4 has an MSM label. This inference function can be helpful in identifying MSM population who are exposed to increased risk of sexual transmitted diseases[34] and for healthcare providers to implement better resource relocation. Further, in the Figure 4 *case b*, the n-ary relationship is addressed. As detailed in Figure 5 example *b*, the answer "had about 3 male sex partners in the past year" cannot be described in a single chain of triples because it would be transcribed as "patient_b" "has_number_of_sex_partner" "3" and "3" "PACO: hasObservationPeriod" "past_year." The latter triple is semantically wrong when represented in natural language. Therefore, we followed our n-ary primary solution and transformed this n-ary relationship into two chains of triples. Then the relationship would be transcribed as "patient_b" "has_sexual_behavior_element" "sexual_behavior_element," "sexual_behavior_element" "has_number_of_sex_partner" "3"; and "sexual_behavior_element" "PACO: hasObservationPeriod" "past_year." In this case, both "number of sex partner" and "observation period" describe a specific sexual behavior of the patient instead of a time that describes a number. In addition, as shown in Figure 5 example *c*, the standardization of numbers can transform the "34 yo" text to float format data "34.0." This can be especially useful in the data transformation and normalization stage in real-world applications. In short, in the two synthetic examples, we displayed how the proposed ontology can function with versatility and help with the modified real-world problems for privacy reasons.

[Insert Figure 5 about here]

**Evaluation Results**

***Semantic Evaluation***

      We evaluated the ontology's semantic meaning in three rounds with the Hootation tool, which transforms ontological hierarchies and relations into accurate natural language phrases to facilitate human expert review.[35] The three evaluators read the phrases and decided whether the organization of the proposed ontology was rational. For example, the relation of "`SDoHO#093` ⊑ ∃ `relates.SDoHO#134`" explains the relation between "Support_Healthy_Eating_Pattern_Access" and "Access_to_Healthy_and_Nutritious_Option," and the tool "Hootation"[35] will transform the hierarchy into "every support_healthy_eating_pattern_access is something that relates an access_to_healthy_and_nutritious_option" for the evaluators to review. We then recorded two agreement scores, inter-evaluator and rational agreement. The inter-evaluator agreement score was calculated by the number of harmonized statements by all three evaluators, divided by the total number of statements. The rational agreement was calculated by dividing only the number of harmonized rational statements by the total number of statements. The "Support_Healthy_Eating_Pattern_Access" example was labeled as rational among the evaluators and counted into the rational agreement and the inter-evaluator agreement. The relation example of "every environmental_justice is an element_relevant_to_socio-environmental_neighborhood," however, was labeled as irrational among all evaluators, and this phrase was counted into the inter-evaluator agreement but not in the rational agreement. Corrections were made after each round of review. Classes and relations marked as irrational by any evaluator were discussed and revised iteratively for best agreed-upon achievement. For the ambiguous concepts, we added a description or definition to the ontology.

For hierarchical irrational concepts, we merged, deleted, and added concepts for the best common sensed structure. For irrational naming conventions, we updated the labels to be more precise.

The first round of evaluation reached 0.53 of the rational agreement, and the second round, 0.837. A final rational agreement score of 0.967 was achieved after three rounds of evaluation, with an inter-evaluator agreement of 0.923. There were nine relations that did not achieve rational agreement and could not be further improved. Of the nine, seven were subclass relations, and two were data property relations. Four out of the seven were related to the definition of diet and its subclasses. Due to various definitions in the Unified Medical Language System (UMLS), agreement was not achieved. Three disagreed on their subclass hierarchical structure, and the two object property relations disagreements were on the level of restriction type. As to the disagreement among evaluators and limited sources for further improvement, the rationale agreement score was pushed to 0.967. For the ones that did not reach rational agreement, we kept those that the majority of evaluators voted on in the latest version of the proposed ontology.

### *Coverage Evaluation*

We evaluated the coverage of SDoHO by utilizing clinical notes and survey questionnaires. To maximally represent the real-world SDoH factors, we used two sets of clinical notes, which contained different types of notes across facilities, and one set of survey questionnaires, which aimed to collect consumers' responses on SDoH elements (see the Methods section). The clinical text was unstructured, non-hierarchical, non-standardized, and uncategorized, and the questionnaire had one layer of hierarchy. The challenge was to represent the non-standardized factors by our standardized terms for the downstream usages. Our proposed ontology aimed to address the challenge by creating hierarchical concepts, object properties, and measurable values. As a result, the three text sources reached seven top-level ontology domains and 30 concept-level factors. The Harris County Psychiatric Center (HCPC) clinical notes, Mayo Chronic Pain cohort, and All of Us survey were mapped to 14, 15, and 13 classes from the ontology, respectively (distribution shown in Supplementary Table 1). Each text source aimed at different health concerns so that the ontology coverage varied for the real-world reflection.

***Clinical Note Coverage***

Coverage of the ontology was measured with psychiatric patient clinical notes. The notes covered mainly six areas: "Demographic," "Element_Relevant_to_Behavior_and_Lifestyle," "Element_Relevant_to_Economic_Stability," "Element_Relevant_to_Education," "Element_Relevant_to_Neighborhood," and "Element_Relevant_to_Social_and_Community_Context." Within the 300 notes, there were 414 SDoH factors mapped to a total of 14 SDoH classes. Among the 14 mapped concepts, the top three factors were "Education_Level," "Living_Status," and "Adverse_Childhood_Experience." As shown in Table 2, the proposed ontology is applicable to fully cover SDoH domains, concepts, and the downstream object properties and measurements in the clinical notes context.

[Insert Table 2 about here]

Another set of clinical notes was the chronic pain cohort dataset, which covers the same six domains but varies in the main concept level from the psychological assessment set (distribution shown in Supplementary Table 1). The top three identified SDoH factors in the chronic pain set were "Substance_Abuse," "Employment_Status," and "Marital_Status." All of the SDoH domains and main concepts from the 507 chronic pain notes were covered by our proposed ontology. In the value and measurement perspective, SDoHO reached 66.67% in the coverage, as summarized in Table 2. The yes-or-no type of measurements or values are fully covered by the ontology. For example, the values for "Insurance" are yes and no. If the patient has insurance, the relation can be "patient" "has_insurance" "insurance." The negative answers can be recorded by a negation assertion, such as "patient" (negative object property assertion ("has_insurance" "insurance"). The proposed ontology also covers some categorical measurements and values. For example, we have categorical permissible values, "good_adherence" and "poor_adherence," for the measure "Adherence." Thus, as shown in Table 2, the coverage of the proposed ontology in the chronic pain cohort reached 100%, 100%, and 66.67% for domain level, main concept level, and value/measurement level, respectively.

### Survey Coverage

The coverage also was evaluated with the SDoH survey from the "All of Us" research program, mainly at the domain level. All domains in the survey were mapped in the ontology. The concept may have different naming representations. For example, the "Loneliness" domain in the survey was presented as "Social_Isolation" in the ontology. Out of the 13 domains with 81 items in the survey, our ontology achieved 100% coverage at the survey's domain level, as seen in Table 2. The main concept-level items were overly granular and specific and were out of the scope of this ontology. For example, in the AoU survey, the domain of item, "people around here are willing to help their neighbors," is "Neighborhood Cohesion," which can be mapped to SDoHO's "Social Integration"; but the item's granularity is too specific to be semantically represented and is, therefore, out of the scope of SDoHO and not included or evaluated at the current stage. In the measurement and value level, 44.44% of the detailed measurements from the survey were covered by the ontology. The classification of frequency measures were included in the ontology. For example, frequency descriptions, such as "Almost_Everyday" or "At_Least_Once_A_Week," were updated into SDoHO. These categories allow the ontology to reflect the activity frequencies from the real world in a quantifiable means. The non-quantifiable measures, such as "very well" or open-question answers, are out of the scope of this ontology. In addition, the level of granularity differs, and some specific measures were not included after consideration. They can be added to enrich the ontology in the future stages. Therefore, the AoU survey is fully covered in the domain level but less than half at the measurements and values level.

**Discussion**

*Contributions*

SDoHO has great potential to fill the gap with standardized and comprehensive SDoH semantic representations. Currently, there are heterogeneity gaps and lack of standardization in the SDoH domain.[20,36,37] Arons et al. found that several concepts related to SDoH, such as incarceration and veteran status, did not have a standard terminology code. Further, Resnick et al. matched standard vocabularies with the Assessing Circumstances & Offering Resources for Needs (ACORN) survey and found the need in SDoH terminology representation.[38] SDoHO aims to provide a formal and standard representation of the domain that could fill in these gaps. Not only does the SDoHO exhaustively cover the concept definitions with standard code, such as UMLS, but it also provides the hierarchical relations and gathered measurements for future applications. Thus, SDoHO can be a standard framework to address an urgent need.

In all, SDoHO has well-designed hierarchies, practical objective properties, and versatile functionalities, and the comprehensive semantic and coverage evaluation achieved promising performance compared to the existing ontologies relevant to SDoH. The aim of the proposed ontology was to provide a comprehensive and formally defined collection of SDoH concepts, hierarchies, and relations that can be adopted by medical and public health settings to ensure data and semantic interoperability. Concepts defined and standardized by UMLS with the Concept Unique Identifier (CUI) and official websites maximally ensured unity and future interoperability. The measurement among concepts and linkage to relevant measuring items enables our proposed ontology to be applicable to clinical, public health, and biomedical informatics perspectives. For data presentation, the ontology will ensure semantic interoperability and data FAIRness (Findable, Accessible, Interoperable, and Reusable). For survey purposes, it will enable standard representation of the input data. In addition, the alignment to clinical notes and the All of Us SDoH survey extends its practicability. By taking into consideration data storage scenarios, SDoHO was designed to be pragmatic. We use identification numbers for the extendable nodes to differentiate the following instances. This design exists in our ontology to differentiate each specific event because they can be followed by further details, such as smoking intensities or frequencies. Gender or race (see Figure 5), however, cannot be further explained or differentiated, so they were not followed by identification numbers. Thus, our proposed ontology was built with a concrete framework design and flexible functions. We also contemplated the potential downstream applications. In leveraging SDoHO with informatics, two possible downstream usages were abstracted, as seen in Figure 6: Natural Language Processing (NLP), to improve the accuracy in identification of SDoH factors from unstructured text and empowering of the computer's understanding of the SDoH factors measurements to assist with clinical decision support.

[Insert Figure 6 about here]

*Limitations and Future Efforts*

In its current stage, the ontology has several limitations and, thus, should be further developed. First, we defined the relations based on current sources, which lack proper hierarchies and relations; therefore, the relationships at their current stage are not fully expressed in the ontology. In response, we will leverage the literature as the main source for relationship extension in the future. Second, the current calculation is annotated in natural language, and queries are not developed in the current version. We thus will utilize Semantic Web Rule Language (SWRL)  to automate the calculation query function. For instance, we can automate the calculation of the healthy food accessibility, using the RFEI score based on geographic location. Instead of using manual calculation, we will link the relevant database and determine the estimation score with the calculation query function. Consequently, our ontology lacks instances and value sets. As such, we will evaluate existing ontologies and published guidelines for the enrichment of the proposed ontology. Currently, SDoHO is aligned with standard terminologies, such as UMLS,[39] LOINC,[31] and Systematized Nomenclature of Medicine-Clinical Terms (SNOMED-CT).[39] For specific domains that cannot be aligned, our current top-down approach with SDoHO can be leveraged as the backbone framework and utilize a bottom-up approach for expansion to align SDoHO with other ontologies, such as those presented in Table 1. Further, we plan to align with Basic Formal Ontology (BFO) to enable interoperability. Finally, as the evaluation results indicated, SDoH information is embedded in text. There is great potential to use the constructed ontology in combination with NLP to volumize the instances of the concepts and further transform the unstructured information to structured. Therefore, our next step is to increase the interoperability of the ontology.

**Methods**

SDoHO was constructed mainly by manual development in Protégé 5.5.0. Figure 6 shows the workflow of the SDoHO designing principles. Data sources and designing considerations are summarized in box *a*. Two evaluation methods, semantic evaluation and coverage evaluation, on the proposed ontology were described in box *b*; and potential future applications are presented in box *c*.

***Data Sources***

To comprehensively incorporate SDoH-related topics, concepts, and knowledge, the data sources that were extensively searched and utilized comprised: (1) multiple official and institutional websites, including WHO,[4] CDC,[40] Healthy People 2020,[41] Healthy People 2030,[2] Kaiser Family Foundation (KFF),[42] Rural Health Information Hub,[43] Healthcare Information and Management Systems Society (HIMSS),[44] NEJM Catalyst,[45] National Academy of Medicine (NAM),[46] Robert Wood Johnson Foundation,[47] and American Hospital Association (AHA)[48]; (2) standardized medical vocabularies and ontologies, including LOINC,[31] SNOMED-CT,[39] and UMLS[39]; (3) relevant biomedical literature[16,36,49,50]; and (4) other useful resources, such as PhenX Toolkit.[30]

***Ontology Design***

The principles for SDoHO design and development are as follows: (1) Interoperability. We extensively reused existing standards and vocabularies to facilitate interoperability between the ontology and surrounding resources. (2) Applicability. One goal of the SDoHO is to support multiple downstream applications. We create classes and properties that offer flexibility to be applied in different contexts, including supporting semantic reasoning and NLP. (3) Scalability. The current version of SDoHO represents meta-level core knowledge of the domain, which is not designed to be exhaustively populated. As efforts accumulate, the ontology will grow and be expanded, leveraging automated informatic techniques. Hence, we will create the classes and properties with optimum hierarchical organization from its initiation, holding space for its scalability and compatibility. In addition, our overall workflow of ontology development can be described as a top-down (knowledge-driven), followed by a bottom-up (data-driven) evaluation/validation and refinement approach.

### *Class Definition*

Most of the SDoH data sources that we evaluated have one to two layers of hierarchy of the topics. After reviewing the categories, we first combined Healthy People 2030's five-domain classification with LOINC's six-domain classification, which has an additional section for food. We also accommodated the definition of SBDH to include the behavioral and lifestyle aspects for comprehensiveness of the non-medical factors. In addition, to extend applicability of the ontology with a societal role and the possible influence for the individual, we added demographic and measurement categories after reviewing OMRSE[21,22] and PhenX Toolkit.[30] We segmented terms that have no further hierarchical structure but a pool of SDoH topics into the nine categories. For example, we divided "Income and Wealth," collected from NAM, into two topics, "Income" and "Wealth," as they can be further developed differently. "Income" and "Element_Relevent_to_Income" can be a superclass of "Annual_Family_Income" and "Poverty," while "Wealth" is a superclass of "Money_and_Resources." For the data sources that involved a layered structure, we maximally obtained their original classification. LOINC has six groups under the subject of SDoH, and each group has related measures or questionnaires. We summarized each unique item and classified it under the corresponding structure into the ontology. Further, LOINC is one layered structure. For the SDoHO, we grouped similar items and segmented items into different levels to create a more accurate definition to reflect the hierarchical information among the classes. The same processes were applied to the PhenX Toolkit and other one- to two-layer data sources. Concepts were primarily defined by aligning with available definitions from the UMLS with the CUI.

### Property Definition

We designed the relationships meticulously in the proposed SDoHO, including object properties, data properties, and annotation properties. We also imported and organically reused some existing ontologies to extend the coverage and flexibility of SDoHO's relationships. We added object properties across domains into the proposed ontology when the relation between classes was identified. Some object properties were reused to formally define different classes when needed. Ranges and domains were restricted for specific object properties. Data properties were created to accommodate the usage between classes and data formats. We also added needed annotation properties in use to further disambiguate concepts.

### Evaluation Methods

#### Semantic evaluation

We first evaluated the ontology semantics. The semantic representation of the concepts and properties, including axioms, subclass hierarchies, and other restrictions from the ontology, were first transformed to natural language sentences using an ontology evaluation tool, Hootation,[35] and evaluated by three human experts. Disagreements on concept and hierarchical definitions were addressed after evaluation. The evaluation processes were repeated three times until no further disagreement could be resolved. Classes that did not achieve rational agreement and that were unjustifiable were summarized.

*Coverage evaluation*

Two sets of real-world clinical notes were utilized for conceptual coverage evaluation of SDoHO. One was a subset of psychosocial assessment notes from HCPC, collected between January 1, 2007, and October 1, 2017 (over 100,000 patients). Social history sections, which were rich in SDoH-related factors, were extracted from these notes. A total of 300 social history sections were randomly selected and manually annotated for ontology evaluation. Another set comprised 507 clinical notes retrieved from the Mayo Clinic and Olmsted Medical Center (OMC).[51] The cohort consisted of local adult patients with noncancer chronic pain who were receiving health care at the Mayo Clinic and/or the OMC between January 1, 2005, and September 30, 2015. The definition for chronic pain was based on Tian et al.[52] Two groups of evaluators collaboratively worked on the identification of SDoH factors from clinical notes.

In addition, the ontology's coverage was evaluated with the NIH All of Us Research Program's SDoH survey,[53] which was not used in the ontology construction phrase. The survey contains 13 domains with measurement items, gathered from various sources, to evaluate the participants' own perceived feelings, influenced by the social surroundings. Each measurement item is a question collected from relevant sources. In our evaluation process, we compared the coverage of classes at the survey's domain level and value/measurement level.

**Data Availability**

The SDoHO is available at https://sbmi.uth.edu/bsdi/offf.htm.

**Code Availability**

Please state.

## Acknowledgments

This study is supported in part by the following grants: NIA 1RF1AG072799, NHGRI 1RM1HG011558, and NIAID 1U24AI171008. We thank the Observational Health Data Sciences and Informatics (OHDSI) program and All of Us Research Program for their work on Social Determinant of Health topics.

## Author Contributions

Y.D. is the first author. C.T., Y.D., and F.L. conceived and designed the work. Y.D. and F.L. constructed the ontology and drafted the manuscript. Y.D., V.K.K, X.H., S.F., and M.Z conducted the data annotation and ontology evaluation. J.D. and J.W.F. provided technical support. Y.D., F.L., V.K.K., A.T.M., and E.Y. designed the use cases. C.T., H.X., H.L., and X.J. supervised the research and critically revised the manuscript. All the authors gave final approval of the completed manuscript version.

## Competing Interests

The authors declare no competing interests.

Figure captions:

Figure 1: Abstract of the partial conceptual framework of SDoHO

Figure 2: Main nine categories in SDoHO and direct subclasses in Protégé

Figure 3: Ontological representation of how geographical information can help with accessibility of healthy food options in Protégé

Figure 4: Overview of two synthetic patient representations

Figure 5: HIV case patient represented in Protégé

Figure 6: Overview schema of SDoHO construction, evaluation, and future application

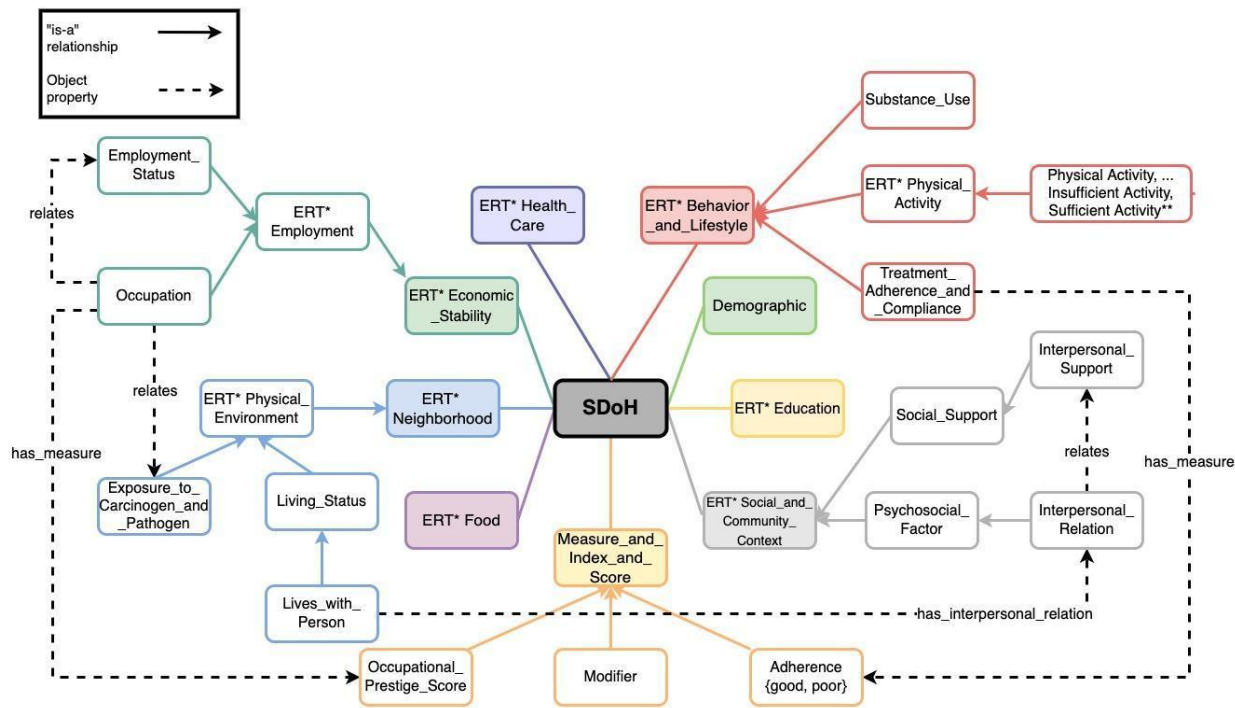

Figure 1: Abstract of the partial conceptual framework of SDoHO

The proposed SDoH ontology has nine main top-level classes under SDoH aspects to represent the relevant factors comprehensively. Sample subclasses are shown to illustrate how SDoH factors were represented and their relationships could be defined using objective properties (simplified in the figure). The figure illustrates the "is-a" relationships by solid lines and the object properties, by dashed lines. The figure displays only partial subclasses and relationships for overview. Data properties and annotation properties are not displayed.

* ERT: Element_Relevant_to_.
** Imported subclasses from PACO.

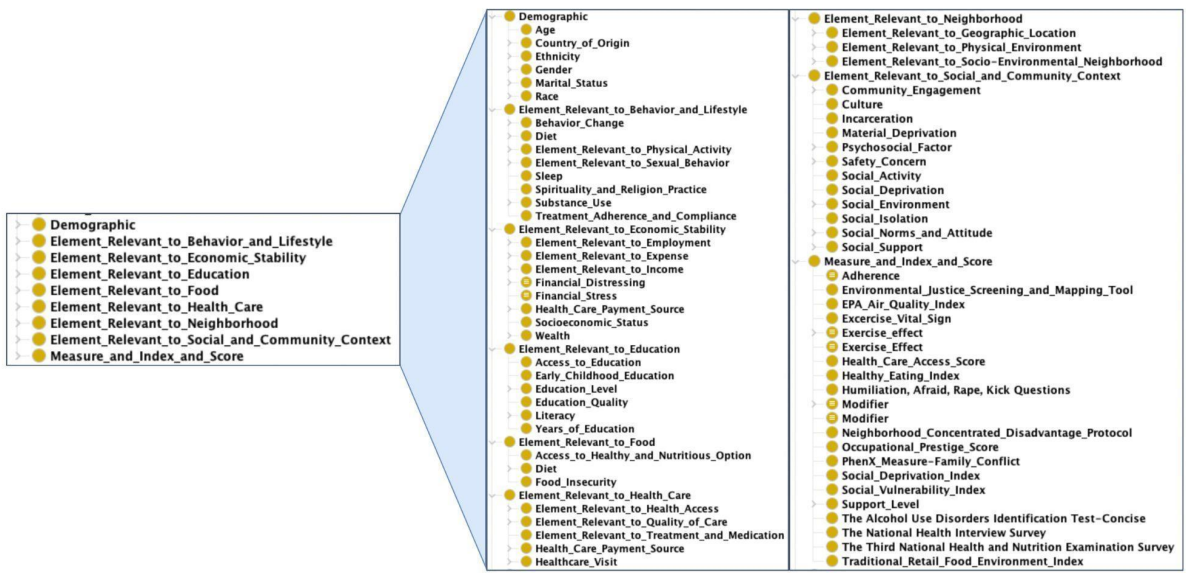

Figure 2: The nine main categories in SDoHO and direct subclasses in Protégé

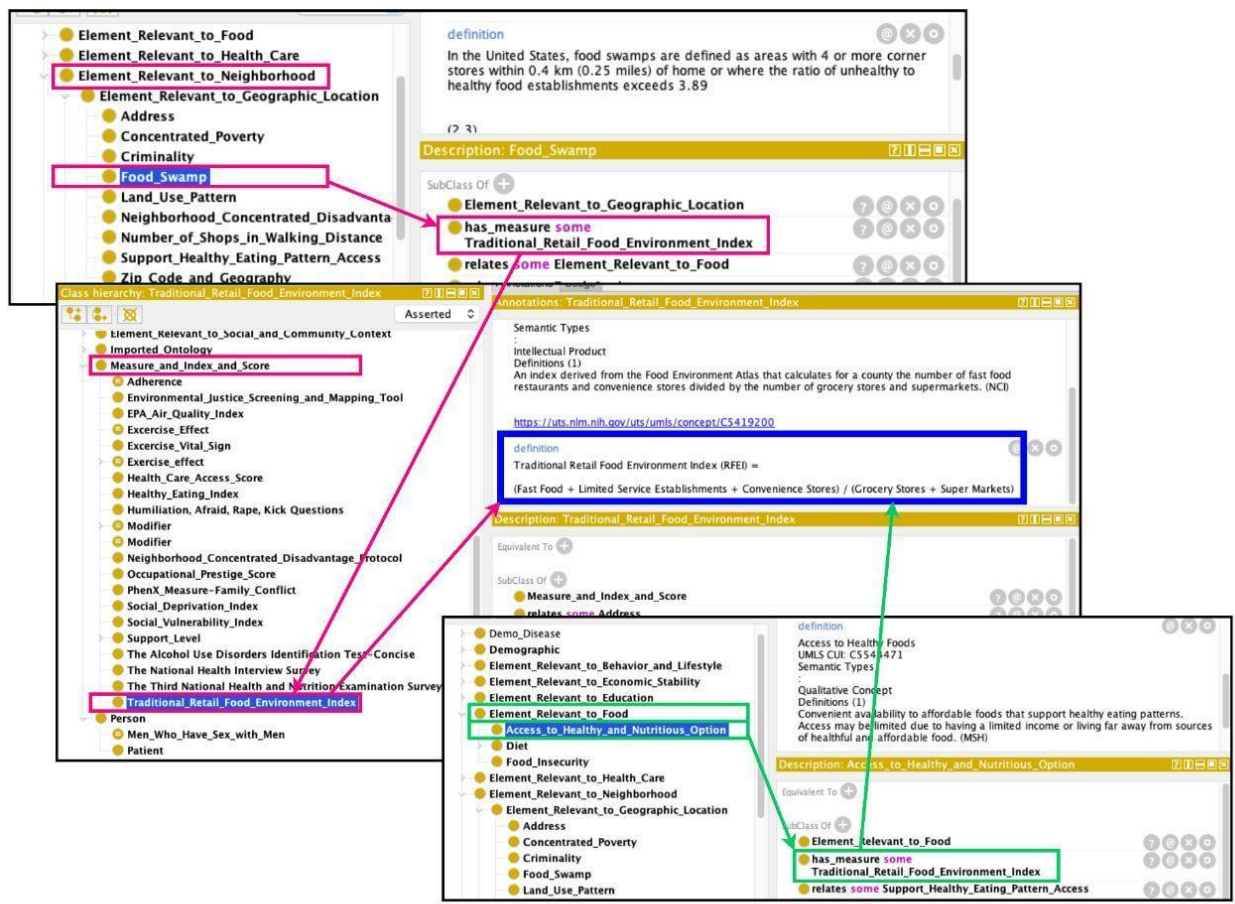

Figure 3: Ontological representation of how geographical information can help with accessibility of healthy food options in Protégé

Example of how to use the "has_measure" object property to represent measures: The concept "Traditional_Retail_Food_Environment_Index" (RFEI) was adopted from the PhenX questionnaire, and the equation involves the availability of healthy food in the target neighborhood. The concept of "Food_Swamp" is defined as "areas with 4 or more corner stores within 0.4 km (0.25 miles) of home or where the ratio of unhealthy to healthy food establishments exceeds 3.89" by Centers for Disease Control and Prevention.[54–56] In this measurement relationship, the score of the RFEI indicates the state of "Food_Swamp" of the neighborhood. Like "Food_Swamp," "Address" and "Zip_Code_and_Geography" are also subdomains of "Element_Relevant_to_Neighborhood" and "Element_Relevant_to_Geographic_Location." In addition, the RFEI score is the measure of the concept "Access_to_Healthy_and_Nutritious_Options."

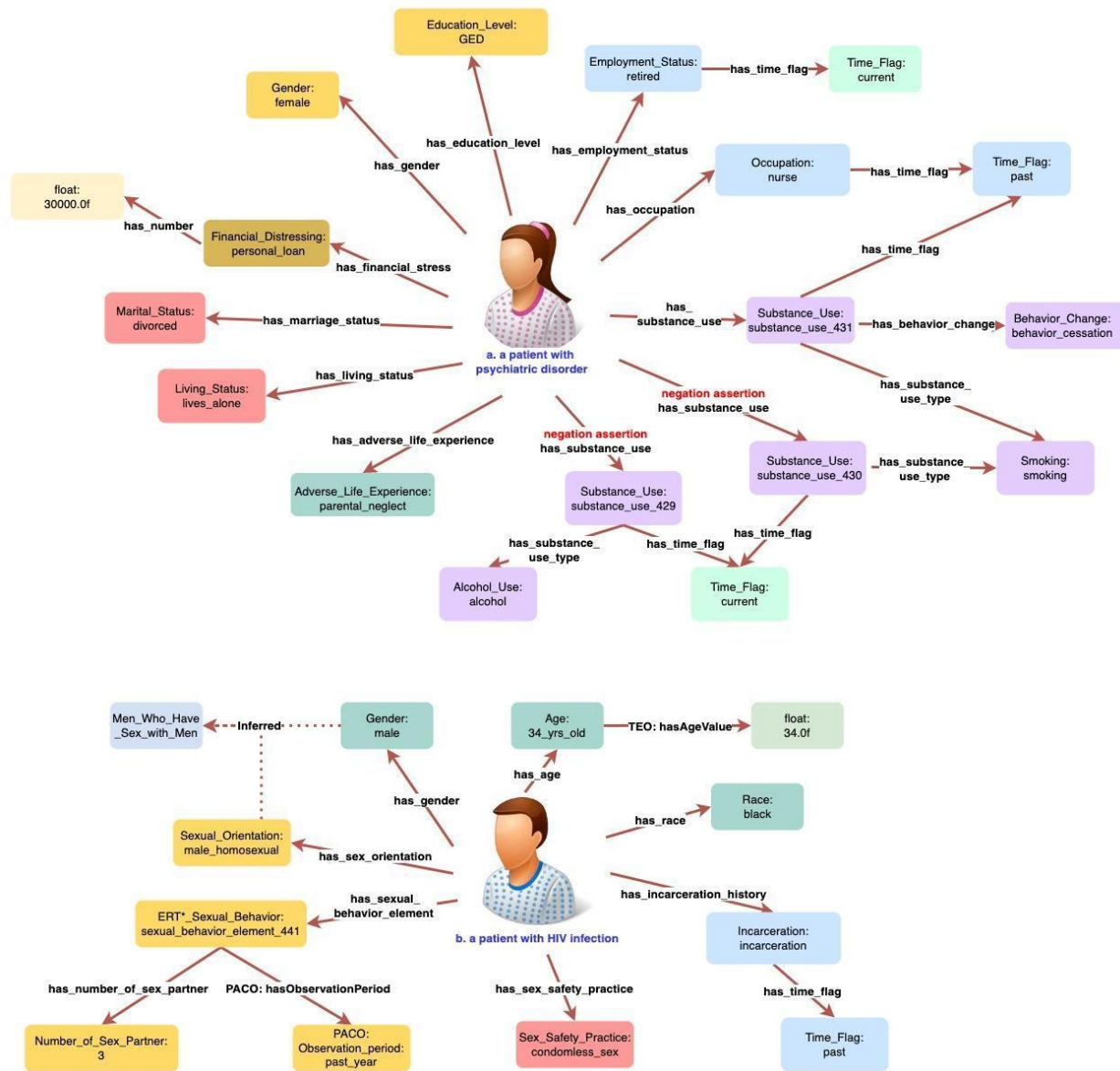

Figure 4: Overview of two synthetic patient representations

Each box color represents essential items extracted from one sentence for a more straightforward review.

a. The data storage representation in the proposed ontology of *case patient a*, who was designed to have a psychiatric disorder. Instances with identification numbers can be further described. The example shows how direct query, time flag, and n-ary relations were represented.

b. The data storage representation of *case patient b*, who was designed to have an HIV infection. The example shows how inference query, other n-ary relations, and numeric normalization issues were addressed with SDoHO.

* ERT: Element_Relevant_to_.

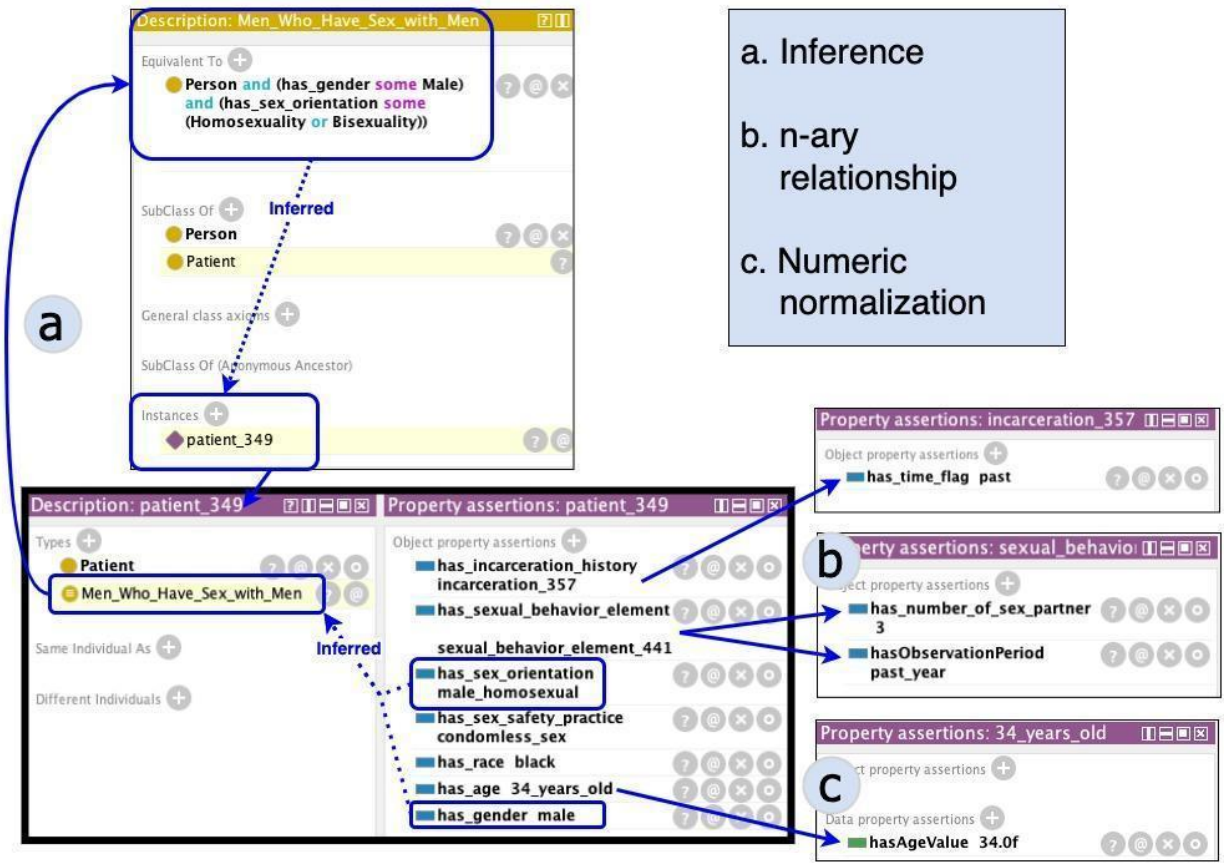

Figure 5: HIV case patient represented in Protégé

Three SDoHO functions for *case patient b* are explained in Protégé.

a. Use of definition in "Equivalent To," inference of the "Men_Who_Have_Sex_with_Men" (MSM) can be labeled for the synthetic *patient b*.
b. The n-ary relationship of issue were addressed.
c. The text data were normalized to float type.

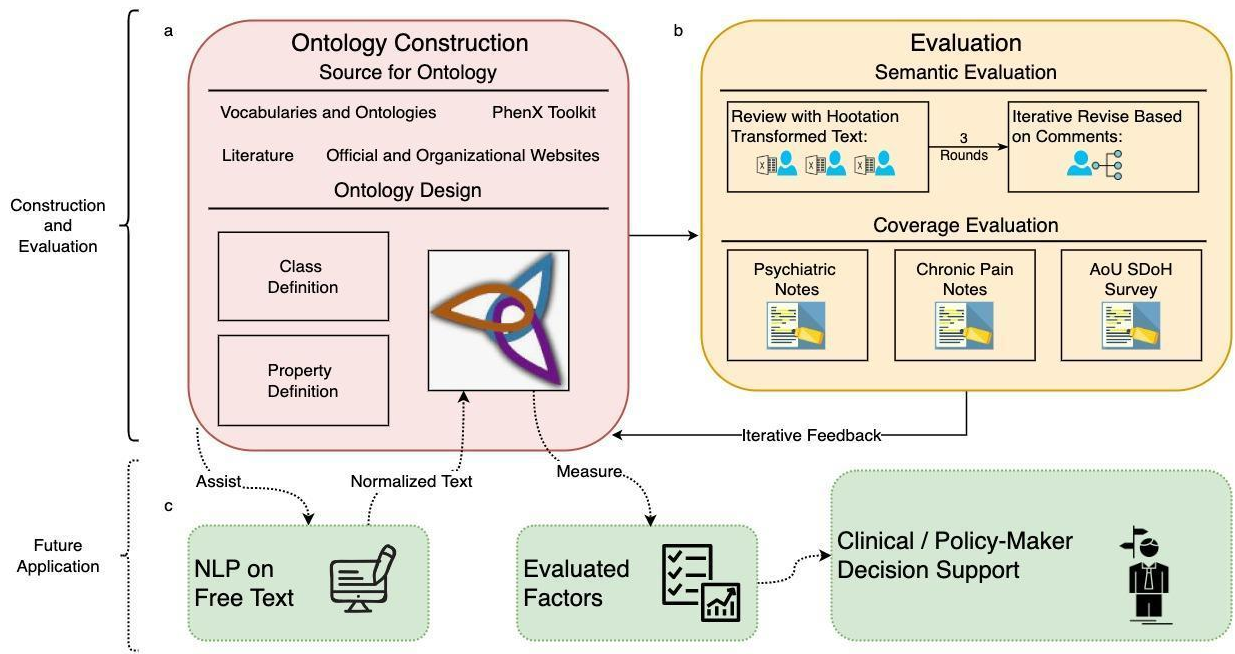

Figure 6: Overview schema of SDoHO construction, evaluation, and future application

a. SDoHO construction, using various sources and designing considerations.
b. Progress of semantic evaluation and coverage evaluation on the SDoHO.
c. Future step of SDoHO application in leveraging NLP tasks and further help with clinical decision support.

Table 1. Comparison of SDoHO and related ontologies.

| Ontology | Description | Representation | Coverage | Lack of coverage | Relationship | Public availability |
|---|---|---|---|---|---|---|
| OMRSE (2011)[21,22] | Developed based on the subjects related to human roles in the society | Formal ontology | Behavior and lifestyle, demographic, education, economics, employment, health care (partially covered) | Food, measurement, neighborhood, most of social and community context | Yes | |
| Public health and clinical social history models (2012)[23] | Conducted from public health surveys and compared to the clinical social history model | Table | Partially covers behavior and lifestyle, substance use/abuse, physical activity, food, economics, employment-occupation; relevant measurement | Demographic, education, health care, neighborhood, social and community context, most areas of economics | None | Yes |
| SMASH (2015)[24,25] | Interrelation among health; weight-related conditions, social activities, and daily physical activities | Formal ontology | Partially covers social context, demographic, behavior and lifestyle, physical activity; calculable measurement | Economics, education, food, health care, neighborhood, most of community context, demographic | Yes | Yes |
| Mobility ontology for PWMD (2018)[26] | Conceptualizing the social and environmental aspects for people with motor disabilities | Formal ontology | Neighborhood; relevant measurement; Demographic, social and community context; partially covers economics, | Behavior and lifestyle, health care | Yes | No |

| | | | | education, food | | |
|---|---|---|---|---|---|---|
| PACO (2019)[27] | Address the heterogeneity of physical activity descriptions | Formal ontology | Partially covers behavior and lifestyle, physical activity; relevant measurement | Demographic, economics, education, food, healthcare, neighborhood, social and community context | Yes | Yes |
| SIREN (2019)[36] | A collection of codes from common medical coding systems related to 20 SDoH factors that involve clinical activities, including screening, diagnosis, and intervention | Excel table | Partially covers economics; covers education, food, healthcare, neighborhood, social and community context | Behavior and lifestyle, demographic, and measurement; most of economics, education, healthcare, neighborhood, social and community context | None | Yes |
| SDoH framework (2022)[28] | An ontology-driven framework covers topics and subtopics from various sources | Framework | Concepts partially covers economics, education, healthcare, neighborhood, social and community context; relevant measurement | No more than one layer of hierarchies | No | No |
| Proposed SDoHO | A formal and standardized representation of the SDoH domain with expandable measurements and relationships | Formal ontology | Demographic, behavior and lifestyle, economics, education, food, health care, measurement, neighborhood, social and | -- | Yes | Yes |

community
context

Table 2. Coverage evaluation results of SDoHO with three textual sources

| Ontology level | Matched in Psychiatry Notes (%) | Matched in Chronic Pain Notes (%) | Matched Items in AoU SDoH Survey (%) |
|---|---|---|---|
| Domain level | 100% | 100% | 100% |
| Main concept level | 100% | 100% | NA |
| Value/measurement level | 100% | 66.67% | 44.44% |

**Supplement**

Table S1. Coverage evaluation results of SDoHO with three textual sources

| SDoH classes | Matched in psychiatry notes (%) | Matched in chronic pain notes (%) | Matched items in AoU SDoH survey (top level) |
|---|---|---|---|
| Demographic | | | |
|   Biological sex | 2 (0.48) | 1 (0.08) | -- |
|   Country of origin | 24 (5.80) | 2 (0.17) | -- |
|   Marital status | 38 (9.18) | 121 (10.09) | -- |
|   Race | 4 (0.97) | 3 (0.25) | -- |
| Behavior and lifestyle | | | |
|   Physical activity | -- | 41 (3.42) | -- |
|   Substance use | -- | 648 (54.05) | -- |
|   Alcohol use | 3 (0.72) | -- | -- |
|   Drug use | 14 (3.38) | -- | -- |
|   Tobacco use | 2 (0.48) | -- | -- |
|   Sexual orientation | -- | 2 (0.17) | -- |
|   Spirituality and religion | -- | -- | 7 |
| Economic stability | | | |
|   Annual family income + financial distress | -- | 3 (0.25) | -- |
|   Employment status | 24 (5.80) | 183 (15.26) | -- |
|   Housing instability | -- | -- | 2 |
|   Insurance | -- | 7 (0.58) | -- |
| Education | | | |
|   Education level | 161 (38.89) | 56 (4.67) | -- |
|   English proficiency | -- | -- | 2 |
| Food | | | |
|   Food insecurity | -- | -- | 2 |
| Neighborhood | | | |
|   Living status | 52 (12.56) | 105 (8.76) | -- |
|   Recreational and leisure opportunities | -- | -- | 8 |
| Social and community context | | | |
|   Adverse childhood experience | 47 (11.35) | -- | -- |
|   Discrimination | -- | -- | 7 |
|   Safety concern | 13 (3.14) | 8 (0.67) | -- |
|   Social disorder | -- | -- | 13 |
|   Social integration | -- | -- | 4 |
|   Social isolation | 19 (4.59) | 9 (0.75) | 8 |
|   Social norms and attitude | -- | -- | 10 |
|   Social support | 11 (2.66) | 10 (0.83) | 8 |
|   Stress | -- | -- | 10 |
| Total count | 414 | 1,199 | 81 |
| Total classes (30) | 14 | 15 | 13 |